\documentstyle[11pt,epsfig]{article} 
\begin{document}


\begin{flushright}
UG-FT-109/99 \\ 
November 1999 \\ 
hep-ph/9911399
\end{flushright}

\renewcommand{\thefootnote}{\fnsymbol{footnote}}
\centerline{\bf Quark mixing: determination of top couplings  
\footnote{Presented at the XXIII International School of
Theoretical Physics ``Recent developments in theory 
of fundamental interactions'', Ustro\'n, Poland, September 15-22,
1999.}
}
\vskip .2cm
\centerline{F. del \'Aguila \footnote{faguila@goliat.ugr.es}}
\vskip .1cm
\centerline{\it Dpto. de F\'{\i}sica Te\'orica y del Cosmos,}
\centerline{\it Universidad de Granada, 18071 Granada, Spain}

\vskip .1cm

{\abstract
The top flavour-changing neutral couplings can be large in 
extended models with vector-like quarks. 
In the next decade(s) the CERN Large Hadron Collider will
allow to measure (bound) them with a precision of 
few per cent.}

PACS {12.15.Mm,12.60.-i, 14.65.Ha, 14.70.-e}
  
\section{Introduction}

The mixing among light quarks $u, d, s, c, b$ is precisely 
measured and in agreement with the Standard Model (SM) ~\cite{SM}.
In contrast the gauge couplings of the top quark are poorly known 
experimentally. However, its mixing is strongly constrained 
in the SM. 
In particular, the Glashow-Iliopoulos-Maiani 
(GIM) mechanism ~\cite{GIM} forbids all tree-level 
flavour-changing neutral currents. On the other hand the top 
mixing can be large in simple SM extensions. 
It is then important to measure (bound) 
it because any positive signal will stand for new physics. 

The top flavour-changing neutral couplings (FCNC) with a light 
quark $q=u,c$ and a $Z$ boson, a photon $A$ or a gluon $G^a$ 
are conveniently \break 
parametrized by the Lagrangian ~\cite{BW}
\begin{eqnarray}
\label{la}
{\cal L} & = & -\frac{g}{2c_W} \bar t \gamma ^\mu \left( 
X^L_{tq} P_L + X^R_{tq} P_R \right) q Z_\mu \nonumber \\ 
& & -\frac {g}{2c_W} \bar t \left( 
\kappa _{tq}^{(1)}-i\kappa _{tq}^{(2)} \gamma _5 \right) 
\frac {i\sigma ^{\mu \nu}q_\nu}{m_t} q Z_\mu \nonumber \\
& & - e \bar t \left( 
\lambda _{tq}^{(1)}-i\lambda _{tq}^{(2)} \gamma _5 \right) 
\frac {i\sigma ^{\mu \nu}q_\nu}{m_t} q A_\mu  \\
& & - g_s \bar t \left( 
\zeta _{tq}^{(1)}-i\zeta _{tq}^{(2)} \gamma _5 \right) 
\frac {i\sigma ^{\mu \nu}q_\nu}{m_t} T^a q G^a_\mu 
 + {\rm h.c.} , \nonumber
\end{eqnarray}
where $P_{L,R} = \frac {1\mp \gamma _5}{2}$ and $T^a$ are the 
Gell-Mann matrices normalized to fulfil 
${\rm Tr} (T^aT^b) = \frac {\delta^{ab}}{2}$. 
These vertices are constrained by the data collected at 
the Fermilab Tevatron ~\cite{TEV}
\begin{equation}
\begin{array}{rcl}
X_{tq} & \equiv & \sqrt{|X^L_{tq}|^2 +|X^R_{tq}|^2} \leq 0.84 ,\\ 
\kappa_{tq} & \equiv & \sqrt{|\kappa^{(1)}_{tq}|^2 +
|\kappa^{(2)}_{tq}|^2} \leq 0.78 ,\\ 
\lambda_{tq} & \equiv & \sqrt{|\lambda^{(1)}_{tq}|^2 +
|\lambda^{(2)}_{tq}|^2} \leq 0.26 ,\\ 
\zeta_{tq} & \equiv & \sqrt{|\zeta^{(1)}_{tq}|^2 +
|\zeta^{(2)}_{tq}|^2} \leq 0.15.  
\end{array}
\label{tevl}
\end{equation}
At this point one can make three obvious questions:
\begin{itemize}
\item How large can these couplings be when all experimental data 
are taken into account ?
\item Do simple models exist saturating the resulting limits ?   
\item How precisely will future colliders measure these vertices ?
\end{itemize}

We will address these three questions in turn. First we review the 
limits on the mixing between light and heavy quarks in Section 2, 
paying special attention to the $t$ couplings. Then in Section 3 
we argue 
that the top mixing can be large in the simplest SM 
extensions with vector-like quarks. Finally 
we discuss the precision with which 
these couplings will be measured at future hadron colliders 
in Section 4. Section 5 is devoted to conclusions. 
For an extended analysis see Ref. ~\cite{JAAS}.


\section{Experimental limits on quark mixing}

Eq. (\ref{la}) is a piece of a general effective Lagrangian 
of which the SM is the lowest order part ~\cite{BW}. All terms 
in $\cal L$ result from dimension six operators after 
spontaneous symmetry breaking. But it can be argued that 
having different origin the size of their coupling constants
can be quite different. The $\gamma ^\mu$ terms are dimension 
four and we expect them to be larger than the $\sigma ^{\mu \nu}$
terms which are dimension five. This is what happens in the 
models we study in next Section. However, when dealing with 
future limits at large colliders in Section 4, we must 
investigate all possible scenarios and then allow for arbitrary 
$\sigma ^{\mu \nu}$ couplings within existing experimental bounds. 
In the rest of this Section we discuss the limits on 
dimension four (renormalizable) couplings. We include charged 
currents because as in the SM, charged (${\cal L}_W$) 
and neutral (${\cal L}_Z$) currents 
are related in SM extensions with vector-like fermions. 

\subsection{Direct limits}

The $W$ couplings $V_{qq'}$ in 
\begin{equation}
{\cal L}_W = -\frac {g}{2\sqrt 2}  \bar q_L V_{qq'} 
\gamma ^\mu q'_L W^+_\mu + {\rm h.c.} ,
\label{lwq}
\end{equation}
where the sum over $q$ and $q'$ 
is understood, are consistent with a $3\times 3$ unitary 
Cabbibo-Kobayashi-Maskawa 
~\cite{CKM} (CKM) matrix ~\cite{PDG},
\begin{equation}
\begin{array}{lll} 
|V_{ud}| = 0.9740 \pm 0.0010, & |V_{us}| = 0.2196 \pm 0.0023, 
& |V_{ub}| = 0.0033 \pm 0.0008,  \cr
|V_{cd}| = 0.224 \pm 0.016, & |V_{cs}| = 1.04 \pm 0.16, 
& |V_{cb}| = 0.0395 \pm 0.0017. 
\end{array}
\label{vckm}
\end{equation}
While the $Z$ couplings $c^{L,R}_{qq'}$ in 
\begin{equation}
{\cal L}_Z = -\frac {g}{2c_W} \left( \bar q_L c^L_{qq'}
\gamma ^\mu q'_L + \bar q_R c^R_{qq'} \gamma ^\mu q'_R \right) Z_\mu , 
\label{lzq}
\end{equation}
where $c^{L,R}_{tq}$ are equal to $X^{L,R}_{tq}$ in Eq. (\ref{la}), 
are diagonal and universal within experimental errors 
~\cite{PDG,FCNC,FAM},
\begin{equation}
\begin{array}{lll} 
|c^{L,R}_{uc}| \leq 1.2\times 10^{-3}, & & \cr
|c^{L,R}_{ds}| \leq 4.1\times 10^{-5}, 
& |c^{L,R}_{db}| \leq 1.1\times 10^{-3}, 
& |c^{L,R}_{sb}| \leq 1.9\times 10^{-3},  
\end{array}
\label{cnd}
\end{equation}
\begin{equation}
\begin{array}{ll} 
c^L_{uu} = 0.656\pm 0.032, & c^L_{cc} = 0.690\pm 0.013, \cr
c^R_{uu} = -0.358\pm 0.026, & c^R_{cc} = -0.321\pm 0.019, \cr
c^L_{dd} = -0.880\pm 0.022, & c^L_{bb} = -0.840\pm 0.005, \cr
c^R_{dd} = -0.054^{+0.154}_{-0.096} , & c^R_{bb} = 0.194\pm 0.018. 
\end{array}
\label{cld}
\end{equation}
The diagonal $u$ and $d$ couplings are measured in atomic parity 
violation and in the SLAC polarized-electron experiments  
and the $c$ and $b$ couplings result from a fit to precise 
electroweak data at the $Z$ peak ~\cite{ZPEAK}. The sign of the 
right-handed (RH) couplings is fixed by the corresponding off-peak 
asymmetries. On the other hand 
the $s$ couplings are less precisely measured but 
also consistent with the SM fermion assignments. The off-diagonal 
couplings in Eq. (\ref{cnd}) vanish in the SM as a consequence of 
the GIM mechanism ~\cite{GIM}. Whereas the diagonal 
couplings in Eq. (\ref{cld}) are a function of a unique angle   
\begin{equation}
c^{L,R}_{qq} = 2 \left( T^{L,R}_{3 q} -Q_q \sin ^2\theta _W \right), 
\label{clrd}
\end{equation}
where $T^{L,R}_{3 q}$ and $Q_q$ are the quark isospin and electric 
charge and $\theta _W$ the electroweak mixing angle. All data in Eqs. 
(\ref{vckm},\ref{cnd},\ref{cld}) agree with the SM within 
experimental errors, but what do we know about the top quark ?

At Fermilab Tevatron it has been established that ~\cite{TAR}
\begin{equation}
\frac {|V_{tb}|^2}{|V_{td}|^2+|V_{ts}|^2+|V_{tb}|^2} = 0.99 \pm 0.29,
\label{rtb}
\end{equation}
however, although consistent with the unitarity of the $3\times 3$ 
CKM matrix, this ratio tells little about it. On the other hand 
(Eq. (\ref{tevl})) 
\begin{equation}
|c^{L,R}_{tq}| = |X^{L,R}_{tq}| \leq 0.84, 
\label{ctq}
\end{equation}
if $t$ mainly decays as in the SM. 
LEP2 data give a similar limit ~\cite{DELPHI}.
Thus, the direct bounds on top mixing, Eqs. 
(\ref{rtb},\ref{ctq}), 
are somewhat weak, leaving room for 
large new effects near the electroweak scale.

\subsection{Indirect limits}

One must also wonder about indirect constraints on quark 
mixing, although they are model dependent. Let us revise 
the estimates discussed in Ref. ~\cite{HPZ} for illustration. 
Assuming only one non-zero FCNC $c^L_{tc}$ in 
Eq. (\ref{lzq}) and 
integrating the heavy top quark the following effective 
Lagrangian for the left-handed (LH) down quarks is 
generated:  
\begin{equation}
{\cal L}^{eff}_Z = - \frac {g}{2c_W}  
\frac {m_t^2}{v^2} \frac {1}{16\pi ^2} ln\frac {\Lambda ^2}{m_t^2}  
\left( V_{cj}^* c^L_{ct}V_{ti} + V_{tj}^* c^L_{tc}V_{ci} \right) 
\bar q_{Lj} \gamma ^\mu q_{Li} Z_\mu + {\rm h.c.},
\label{lef}
\end{equation}
where $\Lambda $ is a cut-off integral and $v$ 
the electroweak vacuum expectation value $\sim 250 \ {\rm GeV}$. 
Then bounds on $c^L_{tc}$ can be derived comparing with neutral 
meson mass differences and rare decays. However, although with a 
minimal set of assumptions, this calculation is too rough and 
the corresponding indirect constraints too stringent. Whatever 
new physics is beyond the SM is unlikely that the net effect is 
only one non-zero FCNC. In general there will be more heavy degrees 
of freedom which after integrating them out will generate new 
effective contributions cancelling partially 
${\cal L}^{eff}_Z$ in Eq. (\ref{lef}).   
This is the case, for instance, in the simplest SM extension 
with a new quark isosinglet $T$ of charge $\frac {2}{3}$. 
As discussed in next Section the 
$V_{cj}^* c^L_{ct}V_{ti} + V_{tj}^* c^L_{tc}V_{ci}$ 
coupling in Eq. (\ref{lef}) must be replaced 
(up to diagonal terms) by 
$V_{\alpha j}^* X_{\alpha \beta}^L V_{\beta i}$, where 
$\alpha , \beta$ run over all quarks of charge $\frac {2}{3}$ 
and $X_{tc}^L = c^L_{tc}$. As 
$X_{\alpha \beta}^L = V_{\alpha m} V_{\beta m}^*$, 
where $m = d, s, b$  is summed up, then  
\begin{equation}
V_{\alpha j}^* X_{\alpha \beta}^L V_{\beta i} = 
V_{\alpha j}^* V_{\alpha m} V_{\beta m}^*V_{\beta i} = 
\delta _{jm}\delta _{mi} = \delta _{ji} ,
\label{vxv}
\end{equation}
implying no FCNC in the down sector if the up quarks are 
degenerate. When the 
actual couplings and masses are introduced, 
the cancellation is not complete.
If $T$ is very heavy, it decouples and $t$ does not have 
large FCNC.

The constraints on RH FCNC appear to be weaker due to the 
absence of RH charged currents in the SM. Still similar 
comments apply. If only one RH coupling 
$c^R_{tc}$ in Eq. (\ref{lzq})
does not vanish, the $\rho $ parameter through the 
$Z$ boson self-energy bounds 
its size. 
Again in definite models this restriction is 
smoothed. If the SM is extended 
with a new heavy quark isodoublet $\left( \begin{array}{c} T \\ B 
\end{array} \right)$ of charges 
$\left( \begin{array}{c} \frac {2}{3} \\ -\frac {1}{3} 
\end{array} \right)$,  
the $T$ and $B$ contributions to the $\rho $ parameter 
cancel if the quarks are degenerate.
If the top mixing is non-negligible the cancellation 
is not complete.

In summary, indirect constraints restrict FCNC but their 
application must be done in a model dependent basis. In fact 
in the two SM extensions just mentioned the top FCNC can be 
large but without saturating the direct bounds in Eq. 
(\ref{ctq}). For example ~\cite{FAM}:  
\begin{equation}
\begin{array}{ll}
|c^L_{tc}| = |X^L_{tc}| \leq  0.082, 
&{\rm for \  an \  extra \  quark \  isosinglet} \ T , \\
|c^R_{tc}| = |X^R_{tc}| \leq  0.16 ,       
&{\rm for \  an \  extra \  quark \  isodoublet}
\ \left( \begin{array}{c} T \\ B \end{array} \right) .
\end{array}
\label{lim}
\end{equation}
The mixing with the $u$ quark can be almost 
as large as with the $c$ 
quark but the top can not have a large mixing with both at 
the same time. Otherwise, it would be also a large coupling 
between $u$ and $c$, what is experimentally excluded 
(Eq. (\ref{cnd})). 


\section{Simple SM extensions with large top mixing}

Let us discuss in more detail the two simplest SM extensions 
with vector-like quarks allowing for a large top mixing 
with the up or charm quark.

\subsection{One extra isosinglet}

The charged and neutral current terms in the Lagrangian read 
(in the current eigenstate basis) 
\begin{equation}
{\cal L}_W = -\frac {g}{2\sqrt 2}  \bar u_{Li}^0 \gamma ^\mu 
d_{Li}^0 W^+_\mu + {\rm h.c.} 
\label{lws}
\end{equation}
and
\begin{equation}
{\cal L}_Z = -\frac {g}{2c_W} \left( \bar u_{Li}^0
\gamma ^\mu u_{Li}^0 - \bar d_{Li}^0
\gamma ^\mu d_{Li}^0 - 2s_W^2 J^{\mu}_{EM} \right) Z_\mu , 
\label{lzs}
\end{equation}
respectively. $i=1,2,3$ are the three standard families. 
Diagonalizing the fermion mass matrices, the current 
eigenstates can be written as linear combinations of 
the mass eigenstates 
\begin{equation}
u_{Li}^0 = U^{uL}_{i\alpha} u_{L\alpha}, \ \  
d_{Li}^0 = U^{dL}_{ij} d_{Lj} .
\label{cme}
\end{equation}
Greek (Latin) indices run always from 1 to 4 (3).
Then 
\begin{equation}
{\cal L}_W = -\frac {g}{2\sqrt 2}  \bar u_{L\alpha} V_{\alpha i} 
\gamma ^\mu d_{Li} W^+_\mu + {\rm h.c.} 
\label{lwsm}
\end{equation}
and
\begin{equation}
{\cal L}_Z = -\frac {g}{2c_W} \left( \bar u_{L\alpha} 
X^{uL}_{\alpha \beta} \gamma ^\mu u_{L\beta} 
- \bar d_{Li} \gamma ^\mu d_{Li} 
- 2s_W^2 J^{\mu}_{EM} \right) Z_\mu , 
\label{lzsm}
\end{equation}
with 
\begin{equation}
V_{\alpha j} = U^{uL *}_{i\alpha}U^{dL}_{ij}
\label{ckms}
\end{equation}
and
\begin{equation}
X^{uL}_{\alpha \beta} = U^{uL *}_{m\alpha} U^{uL}_{m\beta} = 
V_{\alpha m} V^*_{\beta m} = \delta _{\alpha \beta} - 
U^{uL *}_{T^0\alpha} U^{uL}_{T^0\beta} \ ,
\label{ndcs}
\end{equation}
as announced in Section 2 and used in Eq. (\ref{vxv}). 
(Note that we have introduced a superscript to distinguish 
up, $u$, and down, $d$, mixing matrices. We omit it when 
the indices specify the quark flavour, as for instance in 
Eq. (\ref{la}).) 
This type of models have been studied by many authors 
~\cite{FCNC,FAM,VEC}.
The important point here is: how large can the top mixing be ?
$U^{uL}_{T^0\alpha}$ parametrizes the departure from the SM. 
The constraints on $V_{qq'}$ and 
$c^{L,R}_{qq'} = \pm X^{L,R}_{qq'} - 2\delta _{qq'}Q_q \sin ^2\theta _W$ 
(where the $+(-)$ sign is for the up (down) quarks)  
from present data (Eqs. (\ref{vckm},\ref{cnd},\ref{cld})) 
translate into 
($s^{(')}, c^{(')}$ stand for sinus and cosinus of the 
corresponding mixing angle)
\begin{equation}
U^{uL }_{T^0\alpha} = \left(  0 \ \ ss' \ \ sc' \ \ c \right) ,
\label{sndc}
\end{equation}
where the up entry is much smaller to maximize the 
top mixing with the charm quark  
$X^L_{tc} = -s^2s'c'$ in Eq. (\ref{ndcs}), 
obtaining $|X^L_{tc}| \leq 0.082$ ~\cite{FAM}. 
If we want to maximize the top 
mixing with the up quark, the r\^ole of the charm and the 
up is reversed in Eq. (\ref{sndc}), and 
$|X^L_{tu}| \leq 0.047$. The indirect constraints are 
also fulfilled.     

\subsection{One extra isodoublet}

In this model there are RH charged currents and RH FCNC 
which can be large. With an extra quark isodoublet 
$ \left( \begin{array}{c} T^0 \\ B^0 \end{array} \right) $
the charged and neutral current terms in the Lagrangian read
($u^0_{L4}$ ($d^0_{L4}$) corresponds to $T^0_L$ ($B^0_L$))
\begin{equation}
{\cal L}_W = -\frac {g}{2\sqrt 2} \left( \bar u_{L\alpha}^0 
\gamma ^\mu d_{L\alpha}^0 + \bar T_R^0 \gamma ^\mu B_R^0 
\right) W^+_\mu + {\rm h.c.} 
\label{lwd}
\end{equation}
and
\begin{eqnarray}
{\cal L}_Z & = & -\frac {g}{2c_W} ( \ \bar u_{L\alpha}^0
\gamma ^\mu u_{L\alpha}^0 + \bar T_R^0 \gamma ^\mu T_R^0
- \bar d_{L\alpha}^0 \gamma ^\mu d_{L\alpha}^0 \nonumber \\  
& & - \bar B_R^0 \gamma ^\mu B_R^0 
- 2s_W^2 J^{\mu}_{EM}\ )\ Z_\mu , 
\label{lzd}
\end{eqnarray}
in the current eigenstate basis, and
\begin{equation}
{\cal L}_W = -\frac {g}{2\sqrt 2} \left( \bar u_{L\alpha} 
V^L_{\alpha \beta} \gamma ^\mu d_{L\beta} + \bar u_{R\alpha} 
V^R_{\alpha \beta} \gamma ^\mu d_{R\beta}
\right) W^+_\mu + {\rm h.c.} 
\label{lwdm}
\end{equation}
and
\begin{eqnarray}
{\cal L}_Z & = & -\frac {g}{2c_W} ( \ \bar u_{L\alpha} 
\gamma ^\mu u_{L\alpha} + \bar u_{R\alpha} 
X^{uR}_{\alpha \beta} \gamma ^\mu u_{L\beta}
- \bar d_{L\alpha} \gamma ^\mu d_{L\alpha} \nonumber \\ 
& & - \bar d_{R\alpha} X^{dR}_{\alpha \beta} 
\gamma ^\mu d_{R\beta} 
- 2s_W^2 J^{\mu}_{EM}\ )\ Z_\mu , 
\label{lzdm}
\end{eqnarray}
in the mass eigenstate one. The generalized CKM matrices 
are written
\begin{equation}
V^L_{\alpha \beta} = U^{uL *}_{\rho \alpha} U^{dL }_{\rho \beta}, \ \ 
V^R_{\alpha \beta} = U^{uR *}_{T^0 \alpha} U^{dR}_{B^0 \beta}
\label{ckmd}
\end{equation}
and the neutral couplings
\begin{equation}
X^{uR}_{\alpha \beta} = U^{uR *}_{T^0 \alpha} 
U^{uR}_{T^0 \beta}, \ \  
X^{dR}_{\alpha \beta} = U^{dR *}_{B^0 \alpha} 
U^{dR}_{B^0 \beta} , 
\label{ndcd}
\end{equation}
where the unitary matrices $U^{L(R)}$ 
diagonalize the quark mass matrices. 
A similar analysis as for the isosinglet gives
\begin{equation}
U^{uR }_{T^0\alpha} = \left( 0 \ \ s \ \ c \ \ 0 \right) , \ \ 
U^{uR }_{B^0\alpha} = \left( 0 \ \ \epsilon \ \ 0 \ \ \sqrt {1-\epsilon ^2}
\right) ,  
\label{sndd}
\end{equation}
with $X^R_{tc} = sc$ and $|X^R_{tc}| \leq 0.16$ 
~\cite{FAM}. On the other hand the measured value of 
$b \rightarrow s\gamma $ implies $|\epsilon | 
\leq 0.001$. 
If the top mixing with the up quark is maximized, 
$|X^R_{tu}| \leq 0.14$, fulfilling also the indirect 
constraints.     

Once we have shown that the mixing between the top and the up or charm 
quark can be large, we would also like to know how well can 
it be measured. 


\section{Determination of top mixing at large hadron colliders}

The lifetime of the top is too short 
and then, in contrast with the light quark 
flavours, its mixing can not be precisely measured studying its 
bound states. Thus, the precision with which the top properties 
will be known will depend on the ability to measure them at large 
colliders. The available top factories during the next decade 
will be Tevatron and the Large Hadron Collider 
(LHC) at CERN. Let us discuss in the following how the anomalous 
top couplings can be measured at these hadron machines.

Non-dominant contributions from diagonal top vertices will be 
difficult to disentangle, not so the non-diagonal ones which 
can manifest in new production and decay processes. The analysis 
depends on the energy and luminosity. The variation of the parton 
distributions when the energy increases modifies their  
flux and then the relevance of the processes to be considered. 
On the other hand their relative statistical significance can change 
with the luminosity of the collider. 
Unless otherwise stated, our results will 
correspond to LHC which will also provide the highest precision.
 
\subsection{Production processes}

\begin{figure}[htb]
\begin{center}
\epsfig{figure=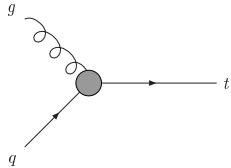, width=3cm}
\vspace*{0.5cm}
\caption{Strong anomalous production of single top at hadron colliders.
\label{fig:2}}
\end{center}
\end{figure}
\begin{figure}[htb]
\begin{center}
\begin{tabular}{ccccc}
\raisebox{-0.8cm}{\epsfig{figure=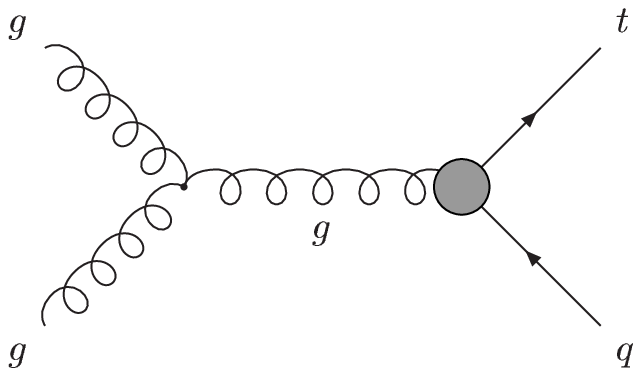, width=3cm}} & + &
\raisebox{-0.8cm}{\epsfig{figure=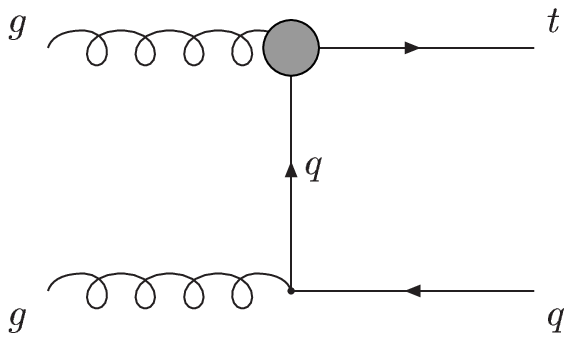, width=3cm}} & + &
\raisebox{-0.8cm}{\epsfig{figure=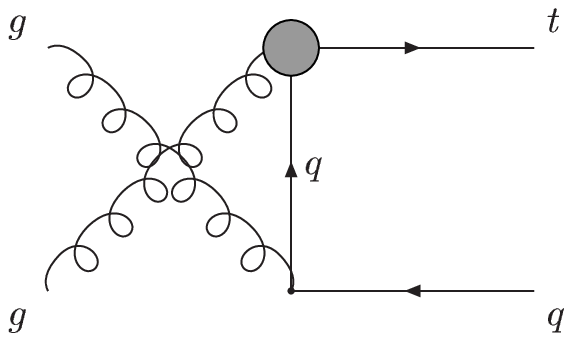, width=3cm}} 
\end{tabular}

\vspace*{0.5cm}
\begin{tabular}{cccc}
+ &
\raisebox{-0.8cm}{\epsfig{figure=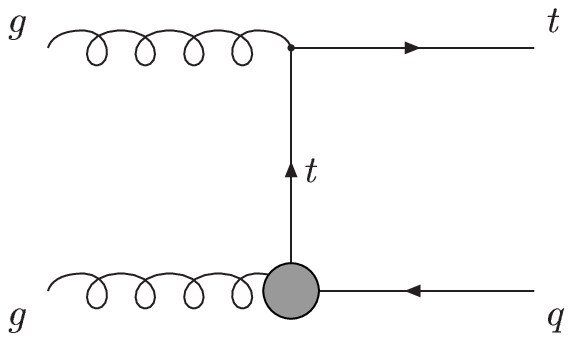, width=3cm}} & + &
\raisebox{-0.8cm}{\epsfig{figure=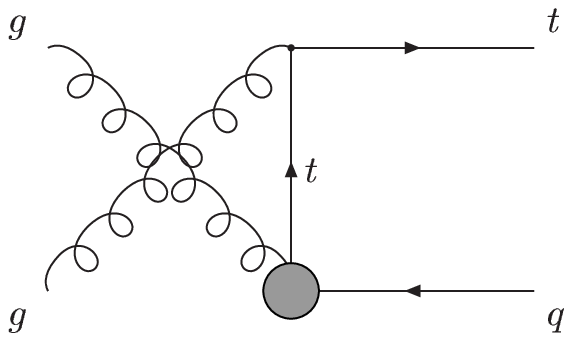, width=3cm}} 
\end{tabular}
\vspace*{0.5cm}
\caption{Strong anomalous production of single top plus a jet 
at hadron colliders.
\label{fig:3}}
\end{center}
\end{figure}

\begin{figure}[htb]
\begin{center}
\begin{tabular}{ccc}
\raisebox{-1.1cm}{\epsfig{figure=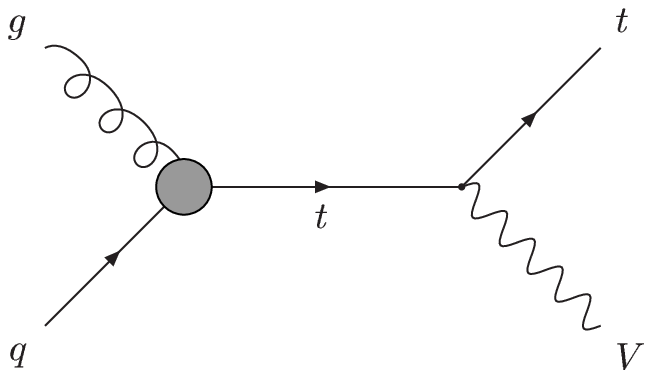, width=4cm}} & + &
\raisebox{-1.1cm}{\epsfig{figure=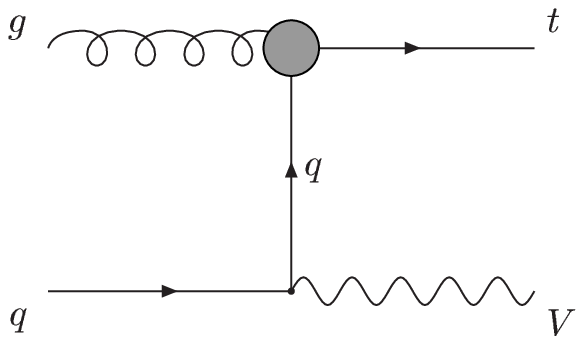, width=4cm}} 
\end{tabular}
\vspace*{0.5cm}
\caption{Strong anomalous production of single top plus a $Z$ boson 
or a photon at hadron colliders.
\label{fig:4}}
\end{center}
\end{figure}

The strong FCNC in Eq. (\ref{la}) can produce a single top 
(Fig. (\ref{fig:2})) ~\cite{HWY} 
or a single top plus a jet (Fig. (\ref{fig:3})) ~\cite{TJE}. 
These are the two lowest order strong anomalous processes 
and must then allow for the best determination of 
the new strong vertices. In the second case 
collisions with initial quarks (or with a gluon and a quark) must 
be also summed up. At LHC, however, the probability of colliding 
two gluons is almost an order of magnitude higher than that of two 
quarks. These vertices can also produce events with a single 
top plus a $Z$ boson or a photon (Fig. (\ref{fig:4})) 
~\cite{AAA}. In Table \ref{Tab:1} we gather the bounds 
on the strong anomalous couplings derived from these processes 
if no signal is observed. These limits give also an indication 
of the precision which can be reached in each case.   
\begin{table}
\begin{center}
\begin{tabular}{lccc}
&& $\zeta _{tu}$ & $\zeta _{tc}$ \\
&&&\\
$pp\rightarrow t\rightarrow l\nu b$  &
&  $7 \times 10^{-4}$ & $2 \times 10^{-3}$ \\ 
$pp\rightarrow tj\rightarrow l\nu bj$ &
& $1.5 \times 10^{-3}$ & $3 \times 10^{-3}$ \\ 
$pp\rightarrow Zt\rightarrow l^+l^-l\nu b$ &
& $7 \times 10^{-3}$ & $1.5 \times 10^{-2}$ \\ 
$pp\rightarrow \gamma t\rightarrow \gamma l\nu b$ &
& $3 \times 10^{-3}$ & $7 \times 10^{-3}$
\end{tabular}
\end{center}
\caption{Reachable bounds on the strong anomalous couplings 
(Eqs. (\ref{la},\ref{tevl})) at LHC with an integrated luminosity 
of $10 \ {\rm fb}^{-1}$. The first two lines are divided by $\sqrt 2$ 
to correct for the use of a different statistics.}
\label{Tab:1}
\end{table}
\begin{figure}[htb]
\begin{center}
\begin{tabular}{ccc}
\raisebox{-1.1cm}{\epsfig{figure=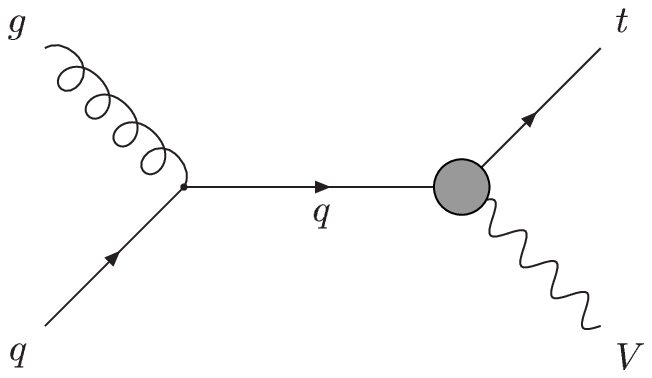, width=4cm}} & + &
\raisebox{-1.1cm}{\epsfig{figure=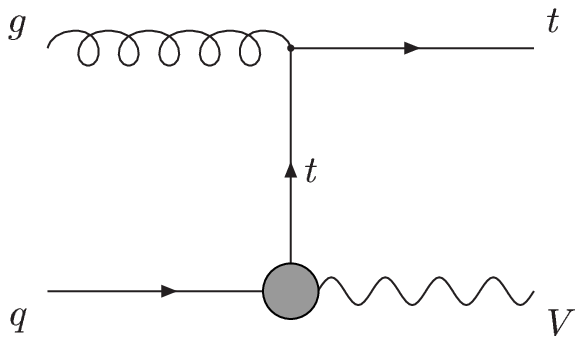, width=4cm}} 
\end{tabular}
\vspace*{0.5cm}
\caption{Electroweak anomalous production of single top plus a 
$Z$ boson or a photon at hadron colliders.
\label{fig:5}}
\end{center}
\end{figure}

The lowest order process involving the weak and electromagnetic 
FCNC in Eq. (\ref{la}) is $Vt$ production, $V=Z, \gamma$, but 
in this case with a standard (anomalous) strong (electroweak) 
vertex (Fig. (\ref{fig:5})) ~\cite{AAA}. 
The top and the $Z$ boson both decay before detection. Then 
we must specify their decay modes to define the sample we are 
interested in. The most significant $Zt \ (\gamma t)$ channel 
is $l^+l^-l\nu b \ (\gamma l\nu b), \ l=e, \mu ,\ $ 
because although it 
has a small branching ratio $ 1.5 \% \ (21.8 \%)$, 
it has also a small 
background $ZWj \ (\gamma Wj)$. 
In Table \ref{Tab:2} we collect the corresponding bounds if 
no signal is observed.   
At Tevatron $\nu \bar \nu jjb$ 
is more significant because in this case the expected 
number of events is smaller and 
this channel has a larger branching 
ratio $13.6 \%$ and the backgrounds $Zjjj, Wt, t\bar t$ 
are not very large. 
\begin{table}
\begin{center}
\begin{tabular}{lccccccccc}
&& $X_{tu}$ & $X_{tc}$ && $\kappa _{tu}$ & $\kappa _{tc}$ & 
& $\lambda _{tu}$ & $\lambda _{tc}$ \\ 
&&&&&&&&&\\ 
$pp\rightarrow Zt\rightarrow l^+l^-l\nu b$ &
& 0.022 & 0.045 && 0.014 & 0.034 &&&\\ 
$pp\rightarrow \gamma t\rightarrow \gamma l\nu b$ &
&&&&&&& 0.005 & 0.013
\end{tabular}
\end{center}
\caption{Most stringent bounds on the electroweak anomalous couplings 
(Eqs. (\ref{la},\ref{tevl})) at LHC with an integrated luminosity 
of $10 \ {\rm fb}^{-1}$.}
\label{Tab:2}
\end{table}

\begin{figure}[p]
\begin{center}
\begin{tabular}{c}
\epsfig{figure=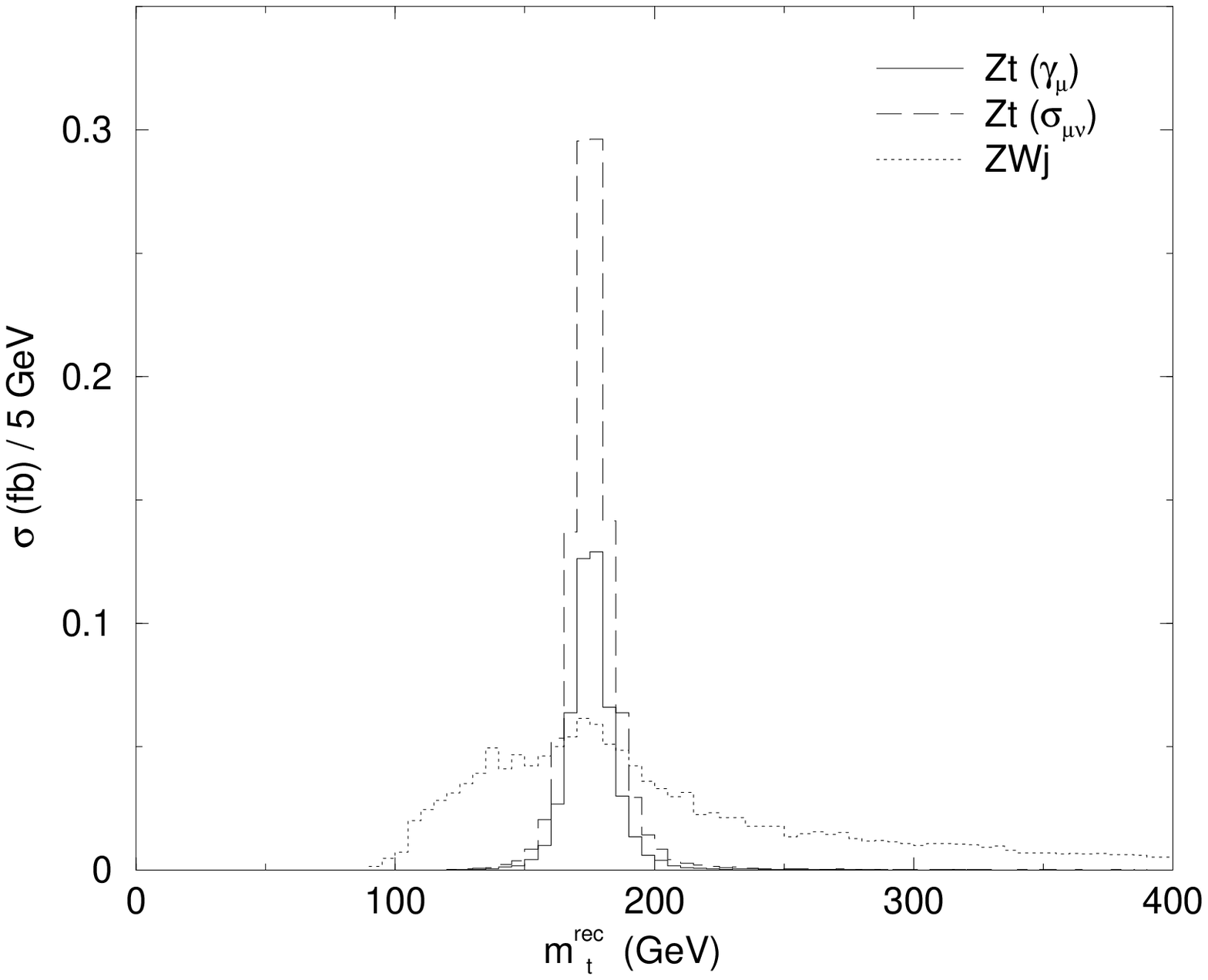, width=6cm} \\
\epsfig{figure=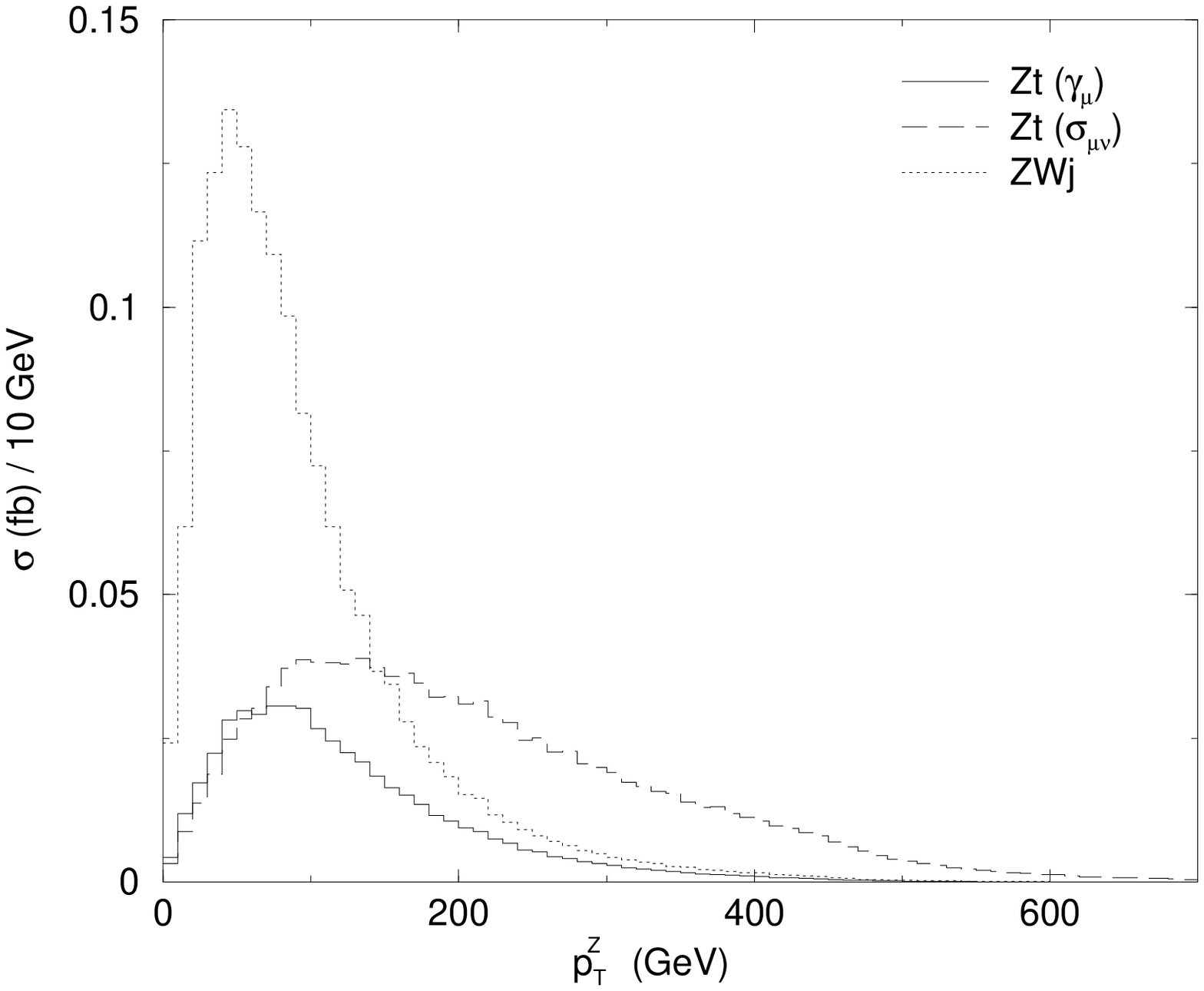, width=6cm} \\
\epsfig{figure=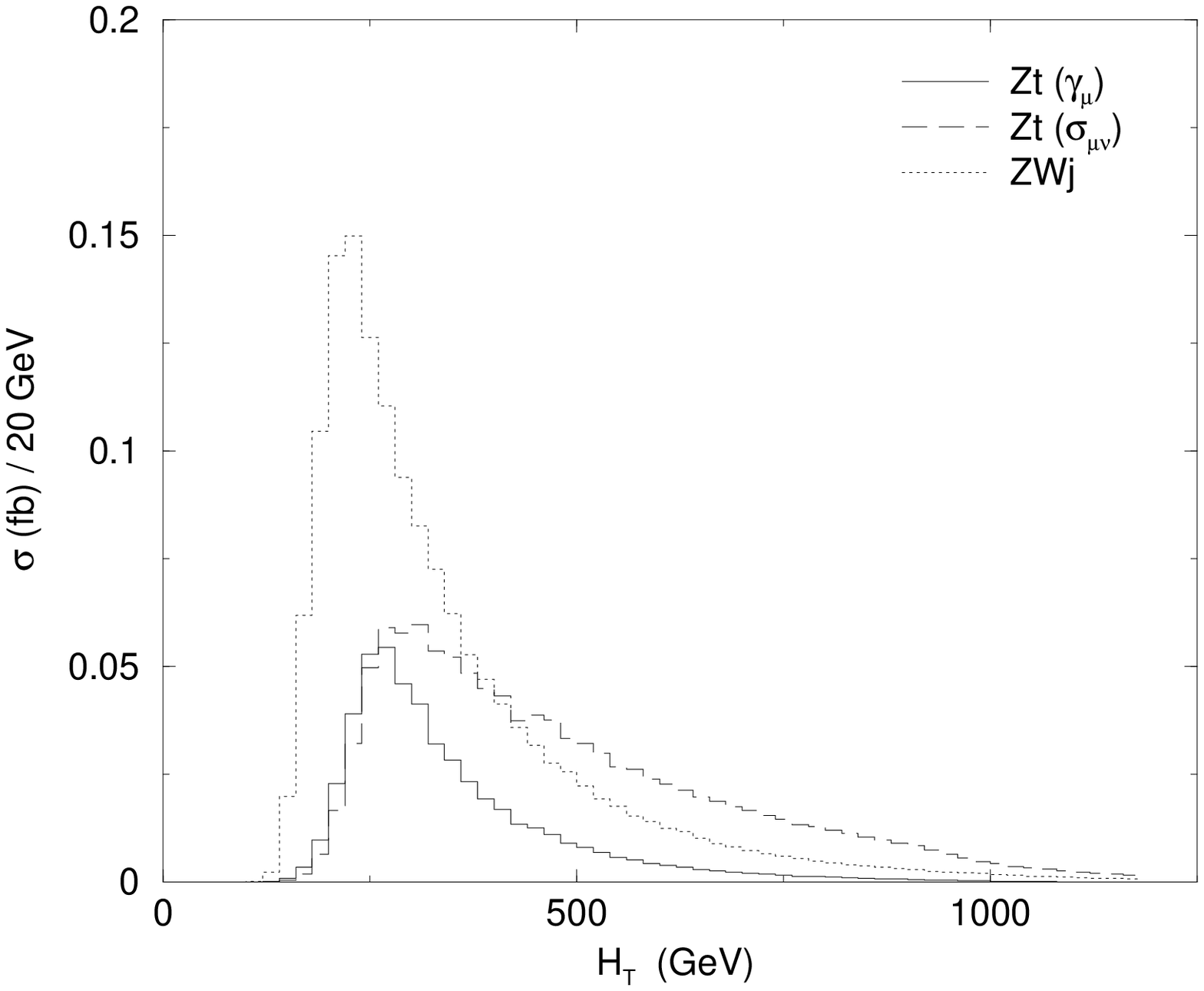, width=6cm} \\
\end{tabular}
\vspace*{0.5cm}
\caption{Reconstructed top mass $m_t^{rec}$, $Z$ transverse momentum 
$p_T^Z$ and total transverse energy $H_T$ distributions before 
kinematical cuts for the $gu\rightarrow l^+l^-l\nu b$ signals 
and background at LHC with an integrated luminosity of $10 \ {\rm fb}^{-1}$. 
$\gamma ^{\mu}$ and $\sigma ^{\mu \nu}$ stand for the 
corresponding non-zero anomalous vertex 
(Eqs. (\ref{la},\ref{tevl})). In this simulation we take 
$X_{tu} = 0.02$ and $\kappa _{tu} = 0.02$ for 
easy comparison. 
\label{fig:6}}
\end{center}
\end{figure}

Let us see now in an example, 
$pp \rightarrow Zt \rightarrow l^+l^-l\nu b$, 
how these limits are derived ~\cite{AAA}.
To estimate the corresponding number of events 
we calculate the exact $2\rightarrow 5\ (gq \rightarrow 
Zt \rightarrow l^+l^-l\nu b)$ squared amplitude and 
use it to generate the events with the correct distribution. 
Afterwards we smear the lepton and jet energies and 
apply the trigger and detector cuts to the resulting 
sample to mimic the experimental set up. 
$b$-tagging is also required. Then we reconstruct 
the $Z \rightarrow l^+l^-$ and $t \rightarrow l\nu b$ 
invariant masses $M^{rec}_Z$ and $m^{rec}_t$, 
respectively (see Fig. (\ref{fig:6})), 
and apply the kinematical cuts 
on $m^{rec}_t$, $p_T^Z$ (the $Z$ tranverse momentum) 
and $H_T$ (the total transverse energy). In this case 
signal and background have the same $M^{rec}_Z$ 
distribution and we do not gain anything cutting on  
this variable. The number of selected events is 
given in Table \ref{Tab:3}. Finally we use the 
adequate statistics ~\cite{FC} 
to derive the expected bounds if no signal is observed. 
\begin{table}
\begin{center}
\begin{tabular}{lcccc}
&& no cuts && with cuts \\ 
&&&&\\ 
$gu\rightarrow Zt\ (\gamma ^{\mu})$ &
& 5.0 && 4.8 \\ 
$gc\rightarrow Zt\ (\gamma ^{\mu})$ &
& 1.1 && 1.1 \\ 
$gu\rightarrow Zt\ (\sigma ^{\mu \nu})$ &
& 11.1 && 10.9 \\ 
$gc\rightarrow Zt\ (\sigma ^{\mu \nu})$ &
& 2.0 && 1.9 \\ 
$ZWq_u$ && 4.9 && 1.4 \\ 
$ZWq_d$ && 5.5 && 1.4 \\ 
$ZWg$ && 4.7 && 1.1
\end{tabular}
\end{center}
\caption{Number of $l^+l^-l\nu b$ events before and after 
kinematical cuts 
($150\ {\rm GeV} \leq m_t^{rec} \leq 200\ {\rm GeV}$ and $200\ 
{\rm GeV} \leq H_T $)
for the $Zt\rightarrow l^+l^-l\nu b$ signals and backgrounds 
at LHC with an integrated luminosity of $10 \ {\rm fb}^{-1}$. 
$\gamma ^{\mu}$ and $\sigma ^{\mu \nu}$ stand for the 
corresponding non-zero anomalous vertex 
(Eqs. (\ref{la},\ref{tevl})). In this simulation we take 
$X_{tq} = 0.02$ and $\kappa _{tq} = 0.02$ for 
easy comparison.}
\label{Tab:3}
\end{table}

\subsection{Decay processes}

\begin{figure}[htb]
\begin{center}
\begin{tabular}{ccc}
\epsfig{figure=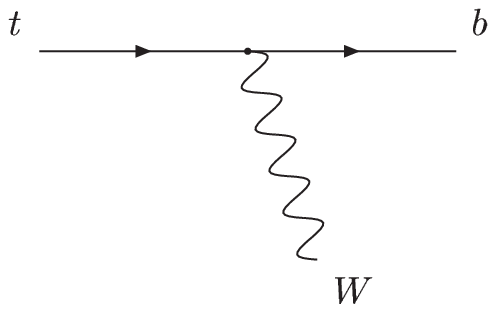, width=4cm} & ~~~ &
\epsfig{figure=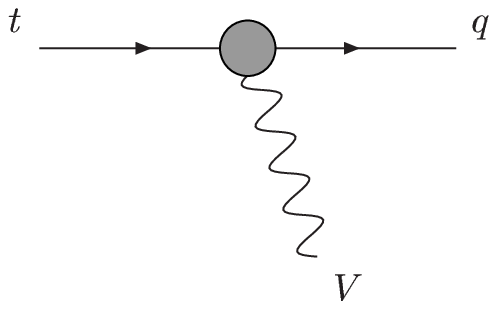, width=4cm} \\[0.2cm]
{\small (a)} & &{\small (b)}
\end{tabular}
\vspace*{0.5cm}
\caption{Standard (a) and anomalous (b) top decay.
\label{fig:7}}
\end{center}
\end{figure}
\begin{figure}[htb]
\begin{center}
\begin{tabular}{ccccc}
\raisebox{-0.8cm}{\epsfig{figure=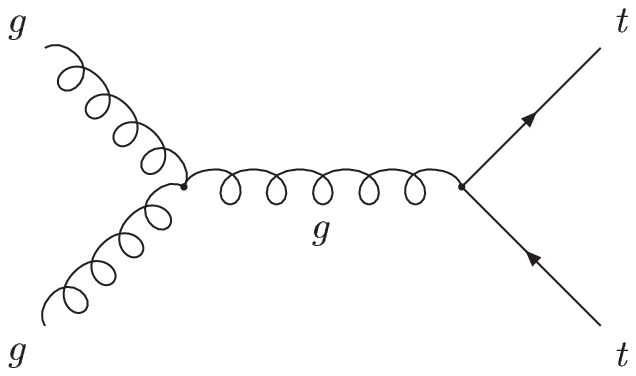, width=3cm}} & + &
\raisebox{-0.8cm}{\epsfig{figure=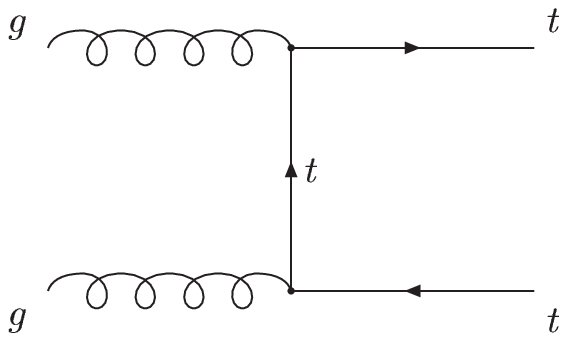, width=3cm}} & + &
\raisebox{-0.8cm}{\epsfig{figure=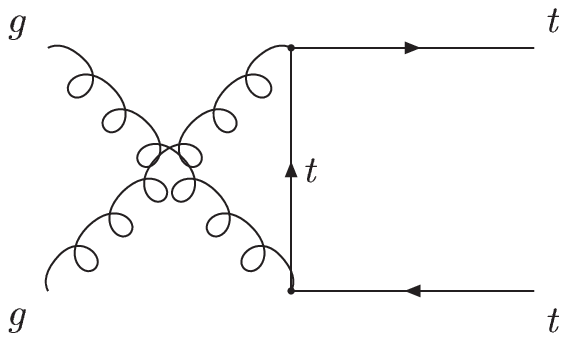, width=3cm}} 
\end{tabular}
\vspace*{0.5cm}
\caption{$gg\rightarrow t\bar t$ production.
\label{fig:8}}
\end{center}
\end{figure}

Up to now we have assumed that the top decays as in the 
SM (diagram (a) in Fig. (\ref{fig:7})), considering only the 
anomalous couplings in the production process. 
This is a good approximation because $t$ decays 
predominantly into $Wb$. However, the large number 
of $t\bar t$ pairs produced at future hadron machines 
(two gluon collisions (Fig. (\ref{fig:8})) stand for 
$90 \%$ of the total cross section and two quark collisions 
for the $10 \%$) allows also to derive competitive bounds 
on top mixing if no anomalous $t$ decay is 
observed (diagram (b) in Fig. (\ref{fig:7})) ~\cite{HPZ,DEC}. 
%


\section{Conclusions}

\begin{figure}[htb]
\begin{center}
\epsfig{figure=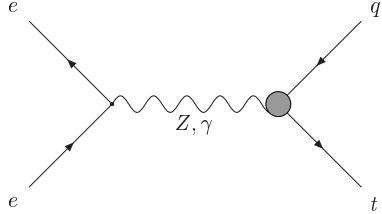, width=5cm}
\vspace*{0.5cm}
\caption{Electroweak anomalous production of single top 
plus a jet at $e^+e^-$ colliders.
\label{fig:9}}
\end{center}
\end{figure}

We have reviewed present direct and indirect limits 
on top mixing, emphasizing that this can be large 
in simple SM extensions with vector-like quarks. 
Future hadron colliders will reduce these bounds 
to the per cent level if no anomalous signal is 
observed. $e^+e^-$ colliders will also set comparable 
limits searching for 
$e^+e^- \rightarrow t \bar q$
(Fig. (\ref{fig:9}))
and anomalous top decays ~\cite{HH}.

\section*{Acknowledgements}
\vspace*{0.5cm}

It is a pleasure to thank J.A. Aguilar-Saavedra for a 
fruitful collaboration when developing this work and 
for help with the drawings, 
and the organizers for their hospitality.
This work was partially supported by CICYT under contract 
AEN96-1672 and by the Junta de Andaluc\'\i a, FQM101. 



\begin{thebibliography}{99}
\bibitem{SM} S.L. Glashow, in Hadrons and their interactions, 
                Proceedings Int. School of Physics `Ettore 
                Majorana', ed. A. Zichichi, Academic Press, 
                New York, 1967; S. Weinberg, Phys. Rev. Lett. 
                {\bf 19}, 1264 (1967); A. Salam, in Elementary 
                particle physics (Nobel Symp. No. 8), ed. 
                N. Svartholm, Almqvist and Wilsell, Stockholm.  
\bibitem{GIM} S.L. Glashow, J. Iliopoulos and L. Maiani, Phys. Rev. 
              {\bf D2}, 1285 (1970).
\bibitem{BW} C. Burgess and H.J. Schnitzer, Nucl. Phys. {\bf B228}, 
             464 (1983);
             C.N. Leung, S.T. Love and S. Rao, Z. Phys. {\bf C31}, 
              433 (1986); W. Buchm\"uller and D. Wyler, Nucl. Phys. 
             {\bf B268}, 621 (1986);
              R.D. Peccei, S. Peris and X. Zhang, Nucl. Phys. 
             {\bf B349}, 305 (1991);
              R. Escribano and E. Mass\'o, Nucl. Phys. {\bf B429}, 
             19 (1994);
              see also S. Bar-Shalom and J. Wudka, hep-ph/9905407 
              and references there in.
\bibitem{TEV} F. Abe {\it et \ al.}, Phys. Rev. Lett. 
              {\bf 80}, 2525 (1998);
              see also T. Han, K. Whisnant, B.-L. Young and X. Zhang, 
              Phys. Lett. {\bf B385}, 311 (1996). 
\bibitem{JAAS} J.A. Aguilar-Saavedra, Ph D Thesis.
\bibitem{CKM} N. Cabibbo, Phys. Rev. Lett. {\bf 10}, 531 (1963);
              M. Kobayashi and T. Maskawa, Prog. Theor. Phys. 
             {\bf 49}, 652 (1973).
\bibitem{PDG} C. Caso {\it et \ al.}, European Phys. Journal 
              {\bf C3}, 1 (1998).
\bibitem{FCNC} Y. Nir and D. Silverman, Phys. Rev. {\bf D42},
                1477 (1990); D. Silverman, Phys. Rev. {\bf D58},
                095006 (1998).
\bibitem{FAM} F. del Aguila, J.A. Aguilar-Saavedra and R. Miquel, 
                Phys. Rev. Lett. {\bf 82}, 1628 (1999);
               F. del Aguila and J.A. Aguilar-Saavedra, 
               hep-ph/9906461.  
\bibitem{ZPEAK} D. Karlen, in Proceedings of the International 
                Conference on High Energy Physics '98, Vancouver, 
                1998.  
\bibitem{TAR} G.F. Tartarelli, CDF Collaboration, in Proceedings 
               of International Europhysics Conference on High 
               Energy Physics (HEP 97), World Scientific, 1997. 
\bibitem{DELPHI} P. Abreu {\it et al.}, Phys. Lett. {\bf B446}, 62 
                (1999); DELPHI note 99-85; DELPHI note 99-146.  
\bibitem{HPZ} T. Han, R.D. Peccei and X. Zhang, Nucl. Phys. {\bf B454},
                527 (1995).
\bibitem{VEC} F. del Aguila and M.J. Bowick, Nucl. Phys. {\bf B224}, 
              107 (1983);
              G.C. Branco and L. Lavoura, Nucl. Phys. {\bf B278}, 
              738 (1986);.
              P. Langacker and D. London, Phys. Rev. {\bf D38}, 
              886 (1988);
              E. Nardi, E. Roulet and D. Tommasini, Nucl. Phys. 
              {\bf B386}, 239 (1992);.
              V. Barger, M.S. Berger and R.J.N. Phillips, 
              Phys. Rev. {\bf D52}, 1663 (1995);
              P.H. Frampton, P.Q. Hung and M. Sher, hep-ph/9903387, Phys. 
              Rep. (in press) and references there in.
\bibitem{HWY} M. Hosch, K. Whisnant and B.-L. Young, Phys. Rev. 
              {\bf D56}, 5725 (1997).
\bibitem{TJE} T. Han, M. Hosch, K. Whisnant, B.-L. Young and X. Zhang, 
              Phys. Rev. {\bf D58}, 073008 (1998).
\bibitem{AAA}  F. del Aguila, J.A. Aguilar-Saavedra and Ll. Ametller, 
                Phys. Lett. {\bf B462}, 310 (1999);
               F. del Aguila and J.A. Aguilar-Saavedra, 
               hep-ph/9909222.
\bibitem{DEC} T. Han, K. Whisnant, B.-L. Young and X. Zhang, 
              Phys. Rev. {\bf D55}, 7241 (1997).
\bibitem{FC} G.J. Feldman and R.D. Cousins, Phys. Rev. {\bf D57}, 
             3873 (1998).
\bibitem{HH} T. Han and J.L Hewett, Phys. Rev. {\bf D60}, 074015 (1999);
             see also V.F. Obraztsov, S.R. Slabospitskii and 
             O.P. Yushchenko, Phys. Lett. {\bf B426}, 393 (1998). 
\end{thebibliography}
\end{document}